\documentclass[12pt]{article}

\usepackage{scicite}
\usepackage{bm}
\usepackage{graphicx}
\usepackage{times}

\newenvironment{sciabstract}{%
\begin{quote} \bf}
{\end{quote}}

\title{Emergent colloidal edge currents generated via exchange dynamics in a broken dimer state}

\author
{Helena Massana-Cid,$^{1,2,\dag}$ Antonio Ortiz-Ambriz,$^{1,2,\dag}$ \\ Andrej Vilfan,$^{3,4}$ Pietro Tierno,$^{1,2,5,\ast}$  \\
\\
\normalsize{$^{1}$Departament de F\'{i}sica de la Mat\`{e}ria Condensada, Universitat de Barcelona, Barcelona, Spain,}\\
\normalsize{$^{2}$Institut de Nanoci\`{e}ncia i Nanotecnologia, Universitat de Barcelona, Barcelona, Spain,}\\
\normalsize{$^{3}$Max Planck Institute for Dynamics and Self-Organization (MPIDS), 37077 G\"{o}ttingen, Germany,}\\
\normalsize{$^{4}$J. Stefan Institute, Jamova 39, SI-1000 Ljubljana, Slovenia,}\\
\normalsize{$^{5}$Universitat de Barcelona Institute of Complex Systems (UBICS), Barcelona, Spain.}\\
\\
\normalsize{$^\dag$Both authors contributed equally to this work.}
\\
\normalsize{$^\ast$To whom correspondence should be addressed; E-mail:  ptierno@ub.edu.}
}

\date{}

\begin{document}

\baselineskip24pt

\maketitle 

\begin{sciabstract}
Controlling the flow of matter down to micrometer-scale confinement is of central importance in materials and environmental sciences, 
with direct applications in nano-microfluidics, drug delivery and biothechnology.
Currents of microparticles are usually generated with external field gradients of different nature [e.g., electric, 
magnetic, optical,
thermal or chemical ones] which are difficult to control over spatially extended regions and samples. 
Here we demonstrate a general strategy to assemble and transport polarizable microparticles in fluid media through combination of confinement and magnetic dipolar interactions.
We use a homogeneous magnetic modulation to assemble dispersed particles into rotating dimeric state and frustrated binary lattices, and generate collective edge currents which arise from a novel, field-synchronized particle exchange process.
These dynamic states are similar to cyclotron and skipping orbits in electronic and molecular systems, thus paving the way toward understanding and engineering similar processes at different length scales across condensed matter.
\end{sciabstract}

Confinement plays a central role in condensed matter physics ({\it 1\/}) and it directly influences 
the transport properties of many systems, from electron flow in graphene ({\it 2\/}) and heterostructure ({\it 3\/}) devices, to vortices in high-T$_c$ superconductors ({\it 4\/}), 
stochastic processes ({\it 5\/}), polymeric ({\it 6\/}), glassy ({\it 7\/}), and active matter ({\it 8\/}) systems.
In colloidal science, confining a suspension of microscopic particles between two flat surfaces reduces the overall mobility, screens  electrostatics and hydrodynamics, and forces the dispersed particles to interact through excluded volume producing novel form of aggregation ({\it 9\/}). 
Confinement can be also used to measure colloidal interactions ({\it 10\/}), probe the frictional deformations ({\it 11\/}) or shed light on subtle 
effects such as critical Casimir forces ({\it 12\/}). 
The combination of a confining mechanism with a pre-designed interaction potential has been predicted to generate novel colloidal phases and dynamics ({\it 13 - 15\/})
[although no net current], all phenomena impossible
to observe in unconstrained systems. 
Despite the rapid progress in colloidal engineering, the few demonstrations of dispersion-free colloidal transport have been achieved only via specially prepared ferromagnetic substrates ({\it 16,17\/}) requiring complex and often difficult fabrication procedures.   
Beyond such limitation, we show here that a combination of hard wall confinement and magnetic dipolar interactions generates colloidal states and currents that can be carefully controlled in-situ by a homogeneous, time-dependent magnetic field.

Our system consists of an ensemble of paramagnetic colloidal particles with diameter $d=2.8 \rm{\mu m}$, dispersed in water and enclosed within a thin cell of thickness $h<2d$, Figs.1(A-D) ({\it 18\/}).
The system is quasi two-dimensionally confined and two particles can only pass each other by moving along the particle plane (unit vectors $\bm{e}_x,\bm{e}_y$), not jumping along the perpendicular ($\bm{e}_z$) direction. 
The particles are doped with nanoscale iron oxide grains and they are responsive to an external magnetic field, $\bm{B}$. 
When $\bm{B}=0$, they are unmagnetized and perform simple thermal diffusion along the same plane.  
Application of the external field induces a dipole moment $\bm{m}=\pi  \chi d^3 \bm{B}/(6\mu_0)$, where $\chi=0.4$ is the magnetic volume susceptibility and $\mu_0$ is the vacuum permeability.
Pairs of particles $(i,j)$ with moments $\bm{m}_{i,j}=m \bm{e}_{i,j}$ and at distance $r=|\bm{r}_i-\bm{r}_j|$ interact through dipolar forces, with an interaction potential given by, $
U_d(r) = \omega 
\left[\frac{\bm{e}_i \cdot \bm{e}_j}{r^3} - \frac{3(\bm{e}_i \cdot \bm{r})(\bm{e}_j \cdot \bm{r})}{r^5} \right]$, with $\omega = \mu_0 m^2/(4 \pi)$.
This potential is maximally attractive (repulsive) for particles with magnetic moments parallel (perpendicular) to $\bm{r}$.
In an unconstrained system, when the applied field is perpendicular to the particle plane, $U_d$ reduces to an isotropic repulsion between parallel particles in the same plane, $U_d(r) = \omega/r^3$. 
This repulsion maximizes the inter-particle distance, inducing the formation of a  triangular lattice close to the bottom plate. 
Upon increasing $\bm{B}$ the strong dipolar repulsion destabilizes this structure, inducing out-of-plane motion and 
the assembly of particles
into vertical columns (along $\bm{e}_z$) ({\it 20\/}).
The presence
of confinement, however, softens the pair repulsion, avoiding the formation of these columns and leading to a different phase behavior ({\it 21\/}). 
If we consider the particles between two hard walls at a distance $h<2d$,
Eq.~\ref{dipolar} can be rewritten as $U_d(r,z) = \omega \frac{r^2-2z^2}{(r^2+z^2)^{5/2}}$. 
In this geometry the particles repel when the elevation difference between their centers is $\Delta z<d/\sqrt{5}$,
but otherwise they experience a short-range attractive and long-range repulsive potential.
Under a static field, it was shown that this change of the potential gives rise to different equilibrium phases characterized by a triangular, square or labyrinthine type of ordering ({\it 21 \/}).

Instead of a static field, we drive the system out-of-equilibrium by using a time-dependent magnetic field which performs a conical precession around $\bm{e}_z$,
\begin{equation}
\bm{B}=B_0[\cos{\theta}\bm{e}_z+\sin{\theta}(\cos{(2 \pi f t)}\bm{e}_x+\sin{(2 \pi f t)}\bm{e}_y)] 
\label{Precession}
\end{equation} 
where $f$ is the frequency and $\theta$ the precession angle that we keep constant at $26.9^\circ$.
The sequence of images in Figs.1(A-C) illustrates the 
impact of the magnetic precession on the particle aggregation process in a cell of thickness $h = 3.9 \rm{\mu m}$. 
Starting from a disordered, fluid like phase
(Fig.1(A)), the magnetic modulation assembles the dispersed particles into a lattice of rotating colloidal dimers, each performing a localized rotational motion around a common vertical axis, Fig.1(B). 
As shown in Movie S1 in ({\it 18 \/}), 
the dimers behave as precessing tops, and the two particles display different brightness due to the differences in elevation,  see also schematics in Fig.1(D). At low driving freqeuncy, the cyclotron frequency $\nu$ of the spinning dimers is phase locked with the driving field, $\nu= f$. 

This dynamic state may be changed  by either raising $f$ or varying the cell thickness $h$. For $h>4.5 \rm{\mu m}$, the spinning dimers transit from synchronous  to asynchronous rotation (SA) above a critical frequency $f_c$, 
where the viscous torque overcomes the magnetic one
an breaks the phase-locking, Fig.1(E). 
In this situation the driving field rotates faster than the dimers and the phase-lag angle between the dimer director and $\bm{B}$ is no longer constant. The fast precession produces a characteristic ``back-and-forth'' dynamics during each field cycle
and a reduction of the average cyclotron frequency. 
For stronger confinements ($h<4.5\rm{\mu m}$), we observe that when $f$ is higher than a threshold rupture frequency $f_r$, 
there is a transition from the synchronous rotation to dimer rupture (SR)
producing two separate lattices of up and down particles, as shown in Fig.1(C).
Even if subjected to gravitational forces, the up-particles remain stably located close to the upper surface as long as the magnetic field
is applied. Similar to the buckled colloidal monolayer of compressed microgels particles ({\it 9\/}), the up-down state can be mapped to a frustrated Ising antiferromagnet on a triangular lattice ({\it 22\/}). However, our lattice features a much larger inter-particle distance, which results from the softness of the pair potentials.

To predict the dynamic states formed by the magnetic dimers at low density, we develop a physical model based on the balance between the magnetic torque $\tau_m$ required to rotate the dimer (Eq.(2) in ({\it 18\/})) and the viscous one $\tau_v$ which arises from the rotation in the fluid (Eq.(4) in ({\it 18\/})). 
The main result is the rotation frequency function,
\begin{equation}
\nu=\nu_0 \frac{\sin{\theta}}{\sin{\alpha}} \sin{\phi} (\sin{\alpha} \sin{\theta}\cos{\phi}+\cos\alpha \cos\theta) \, \, ,
\label{result}
\end{equation}
where $\alpha$ is the minimum angle between the center to center vector joining two particles and the vertical axis, $\phi$ the angle between the horizontal projections of the magnetic field and the dimer orientation,  
$\nu_0=\frac{\chi^2 B^2 d}{24  \mu_0 \gamma}\sim 14.4 \rm{Hz}$, and $\gamma$ the viscous drag of the particle ({\it 18\/}).  The maximum rotation frequency $f_c$ is determined by the maximal value of $\nu(\phi)$. Above $f_c$, synchronous rotation is not possible and phase-slippage occurs. Alternatively, the dimer can rupture at a lower frequency $f_r$ ({\it 18\/}). The resulting frequencies, $f_c$ and $f_r$, are compared with experimental values in Fig.1(F). Even if this analysis only considers the stability of individual dimers, it captures well the dynamics 
of the ensemble at low densities, due to the large inter-particle distance.

Increasing the particle density 
forces the dimers to interact producing collective states where they can break up and their components exchange position with other particles in nearest dimers. 
Experiments and numerical simulations ({\it 18\/}) are used in tandem to unveil the rich dynamic paths that arise by changing $h$ and the normalized area packing fraction, $\Phi=N\pi d^2/(4A)$, where $N$ is the number of particles and $A$ the observation area.
The colored arrows in Fig.2(A) summarize the four situations encountered, while the diagram in Fig.2(B) shows the good agreement between the experimental data (scattered points) and the numerical simulations (shaded regions).
In all cases, we start from an ensemble of spinning dimers in the synchronous state. 
We find that for packing fraction $\Phi >0.35$ the SA an SR transitions display an intermediate exchange state, which gives rise to the Synchronous-Exchange-Asynchronous (SEA) and the Synchronous-Exchange-Rupture (SER) regimes.
 
We characterize these regimes by measuring 
the average cyclotron frequency, Fig.2(C)
and the average distance $\langle \Delta r \rangle$ between nearest neighbors, Fig.2(D). The first observable shows that 
the SA regime (orange line) can be described by the single dimer 
behavior, fitting the curve with the function $\nu=\nu(f)$
in caption of Fig.1, while the SEA (red line) shows a large deviation after the synchronous state. The other two states, SR and SER disappear quickly after the synchronous state since they show no stable  dimers. Thus, we use the second observable,  shown in Fig.2(D),  to characterize these regimes.
For a system in the SA regime, the average distance has a constant plateau with $\langle \Delta r \rangle\approx d$ since the dimers do not break, but simply slow down their orbital motion.
In contrast, $\langle \Delta r \rangle$ reaches the maximum value for the SR path, when the colloids form the up and down lattice. The SEA and SER states are observed at higher packing fraction which corresponds to a lower separation distance.  
In the SEA regime, $\langle \Delta r \rangle$ momentarily increases above $d$ before falling back when the system return to an asynchronous rotation. While the SR regime shows a quick transition to its maximum distance, the SER displays an intermediate region where the average distance depends linearly with the frequency. 

For an isotropic precessing field, the exchange events occur randomly through the sample and arise from a spontaneous dynamical symmetry breaking process. 
We create a polarized edge current by imposing a small in-plane field $B_s$ that breaks the spatial isotropy of the precession. 
The result of this bias is a synchronized bidirectional flux of particles where the up and down paramagnetic colloids periodically separate and recombine as shown in Fig.3(A,B), Movie S3 in ({\it 18\/}).
We quantify the amplitude of this current by measuring the mean edge velocity $\bm{v}_e$ of a particle $i$ at elevation $z_i$ from the middle plane as, 
$\bm{v}_e = \langle  (v_{y,i} \bm{e}_x - v_{x,i} \bm{e}_y ) z_i /|z_i |\rangle$, Fig.3(C). 
This expression averages the in plane velocity components of all particles, giving the same sign to opposite propelling particles located on different planes.
The edge current normalized to the the maximum particle velocity
is thus given by, $I =2|\bm{v}_e|/(f d \sqrt{\frac{\pi}{\sqrt{3} \Phi}})$. The maximum current $I=1$ corresponds to the fully synchronized transport, where all particles travel one lattice site per field cycle. 
Fig.3(C) shows the edge current versus shift angle $\delta$ for different frequencies. 
For high $f$ (blue lines) the current increases slowly with 
$\delta$; the system is initially in an asynchronous state with few exchange events that increase in probability with higher bias. 
In contrast, at low $f$ (red lines) there is no initial current for low $\delta$ and the system starts from a synchronous state.  Above a critical bias, the events of particle exchange become directional and the system quickly reaches the maximum current.
The emergence of the edge current $I$ can be understood by carefully analyzing the exchange and recombination process of two particles in Figs.3(B,E-F). 
When two approaching up and down particles are pushed close to each other by the repulsion from their neighbors, their dipolar interaction is first repulsive, but it quickly becomes attractive creating a new dimer, (pink region in Figs.3(E,F)). 
However, the viscous torque impedes the dimer to rotate as fast as the field. This effect appears in Fig.3(F) where the in plane angle between the field and the dimer director is not constant, but decreases with time starting from $\phi = \pi$. 
When the relative angle becomes close enough to $\pi/2$ the dipole moments are parallel and the dipolar force becomes repulsive. Now the dimer breaks
and the composing particles are pushed apart, coupling to the next particle of the lattice in a \textit{minuet} type of motion. The result is in a bidirectional edge current of up and down particles that flow along the same direction but in opposite sense.

In conclusion, we have shown that the combination of confinement and time dependent magnetic fields produces
a rich repertoire of dynamic states starting from a population of inert colloidal spheres.
We realize self-assembled dimers with localized orbital motion, frustrated binary lattices and robust bidirectional edge currents.
This transport dynamics resembles the skipping orbits in a dilute electron gas moving through semiconductor heterostructures ({\it 23\/}),
 inviting a fundamental study of the analogy between both process at a semi-classical level.
The exotic states we report here via an engineered soft-shoulder potential were predicted in different theoretical models, but not observed 
so far ({\it 13,15\/}). Our results also open a new avenue in 
manipulation of microscopic matter in fluids, with great potential for non-invasive transport of colloidal matter trough small membranes and pores where confinement plays a key role.   

\begin{quote}
{\bf References and Notes}

\begin{enumerate}
\item L. D. Gelb, K. E. Gubbins, R. Radhakrishnan, M. Sliwinska-Bartkowiak, {\it Rep. Prog. Phys.} \textbf{62}, 1573–1659 (1999).
\item A. H. Castro Neto, F. Guinea, N. M. R. Peres, K. S. Novoselov, A. K. Geim, {\it Rev. Mod. Phys.} \textbf{81}, 109 (2009).
\item C. Gong, X. Zhang, {\it Science} \textbf{363}, eaav4450 (2019).
\item G. Blatter, M. V. Feigel'man, V. B. Geshkenbein, A. I. Larkin, V. M. Vinokur,  {\it  Rev. Mod. Phys.} \textbf{66}, 1125--1388 (1994).
\item T. Gu\'erin, N. Levernier, O. B\'enichou, R. Voituriez, {\it Nature} \textbf{534}, 356-359 (2016).
\item S. Zhu, Y. Liu, M. H. Rafailovich, J. Sokolov, D.  Gersappe, D. A. Winesett, H. Ade, {\it Nature} \textbf{400}, 49-51 (1999).
\item E. Flenner, G. Szamel, {\it Nat. Comm.} \textbf{6}, 392 (2015).
\item A. Bricard, J. B. Caussin, N. Desreumaux, O. Dauchot, D. Bartolo, {\it Nature} \textbf{503}, 95–98 (2013).
\item Y. Han, Y. Shokef, A. M. Alsayed, P. Yunker, T. C. Lubensky, A. G. Yodh, {\it Nature } \textbf{456}, 898 (2008).
\item Y. Liang, N. Hilal, P. Langston, V. Starov
 {\it Adv. Coll. Int. Sci.} \textbf{134–135}, 151–166 (2007).
\item T. Bohlein, J. Mikhael, C. Bechinger, {\it
Nat Mater.} \textbf{11}, 126-30 (2011).
\item C. Hertlein, L. Helden, A. Gambassi, S. Dietrich, C. Bechinger, 
{\it Nature} \textbf{451} 172-5 (2008). 
\item P. J. Camp, {\it Phys. Rev. E} \textbf{68}, 061506 (2003).
\item C. Reichhardt, C. J. Olson, {\it 
Phys. Rev. Lett.} \textbf{88}, 248301 (2002).
\item J. Dobnikar, J. Fornleitner,  G. Kahl, {\it J. Phys. Cond. Matter} 20, 494220 (2008).
\item B. B. Yellen, O. Hovorka, G. Friedman, {\it
Proc. Natl. Acad. Sci. U.S.A.} \textbf{102}, 8860-8864 (2005).   
\item P. Tierno, F. Sagu\'es, T. H. Johansen, T. M. Fischer, {\it
Phys. Chem. Chem. Phys.} \textbf{11}, 9615-25 (2009).
\item See supplementary materials.
\item A. C\ifmmode \bar{e}\else \={e}\fi{}bers, M. Ozols, {\it
Phys. Rev. E}  \textbf{73}, 021505 (2006).
\item Y. H. Miao, D. L. Geng, L. E. Helseth, {\it Langmuir} \textbf{22}, 5572-5574 (2006).
\item N. Osterman, D. Babic, I. Poberaj, J. Dobnikar, P. Ziherl, {\it Phys. Rev. Lett.}  \textbf{99}, 248301 (2007).
\item G. H. Wannier, {\it Phys. Rev. }  \textbf{79}, 357 (1950).
\item C. W. J. Beenakker and H. van Houten and B. J. van Wees,  {\it Superlattices Microstruct.}  \textbf{5}, 127–132 (1989).
\item S. Plimton, {\it J. Comp. Phys.} \textbf{117}, 1-19 (1995).
\end{enumerate}
\end{quote}

\section*{Acknowledgments}
We thank Fernando Martinez-Pedrero for initial experiments,
Thomas M. Fischer and Yair Shokef for stimulating discussions. 
H. M.-C., A. O.-A. and P. T. acknowledge support from the European Research Council, grant agreement No. 811234. 
A. V. acknowledges support from the Slovenian Research Agency (grant P1-0099). 
P. T. acknowledges support from 
MINECO under projects FIS2016-78507-C2-2-P, ERC2018-092827 
and Generalitat de Catalunya under project 2017SGR1061
and programa ``ICREA Acad\`emia''.

\section*{Author contributions}
H. M.-C. performed the experiments. A. O.-A. 
performed the numerical simulations. A. V. 
developed the theoretical model. 
P. T. supervised the work and wrote the paper.
All the authors discussed and interpreted the results.

\section*{Competing interests}
The authors declare no competing interest. 

\section*{Data and materials availability} All data needed to evaluate our conclusions are present in the paper and/or the supplementary materials. 
   
\section*{Supplementary materials}
Materials and Methods\\
Captions for Movies S1 to S4\\
Movies S1 to S4\\
Reference \textit{(24)}

\clearpage

\begin{figure*}[htbp]
\begin{center}
\includegraphics[width=\textwidth]{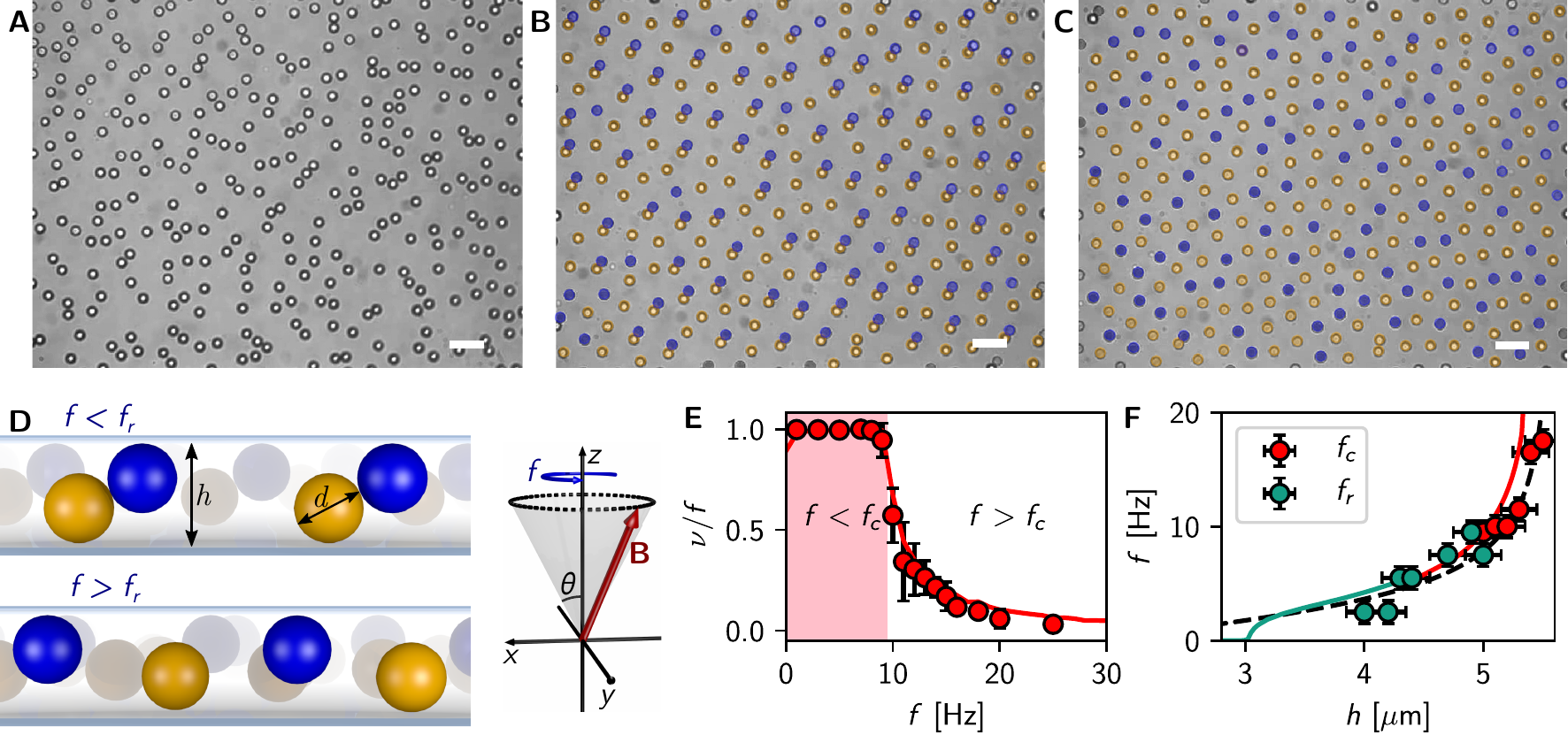}
\caption{\label{fig1}\textsf{\textbf{Assembly procedure of colloidal dimers and binary lattices.} (\textbf{A-C}) Microscope snapshots showing $N=288$ particles in the absence of an external field \textbf{(A)}, and driven by a precessing field with $B_0=7.3 \rm{mT}$, $\theta= 26.9^{\circ}$ 
and frequency $f= 1\rm{Hz}$  (\textbf{b}) and $f= 20\rm{Hz}$ (\textbf{C}).
The dispersed particles (\textbf{A}) are first assembled to rotating dimers (\textbf{B}), and later broken into a binary arrangement of up and down particles (\textbf{C}) by raising $f$. For clarity, particles located close to the upper (lower) plate are shaded in blue (orange), the cell thickness is $h=3.9\rm{\mu m}$. Scale bars are $10 \rm{\mu m}$ for all images, Movie S1 in ({\it 18\/}).
(\textbf{D}) Schematic showing the particle location inside a thin cell ($d<h<2d$) when forming the dimers (top) and the up-down state (bottom). 
(\textbf{E,F}) Dynamics of single dimer: (\textbf{E}) cyclotron frequency $\nu$ of a dimer {\it vs} driving frequency $f$ showing the transition from synchronous to asynchronous regime 
at $f_c=9.8 \rm{Hz}$ from experiments (scattered data) and simulation (continuous line). In the synchronous phase $\nu = f$, while above $f_c$ the cyclotron rotation decreases following 
the equation $\nu=f - \sqrt{f^2-f_c^2}$  ({\it 19\/}). 
(\textbf{F}) Driving frequency versus confinement showing the transition of the dimers from synchronous to asynchronous (red, threshold at $f_c$), or from synchronous to rupture (green, threshold at $f_r$).
Experimental points are scattered data, theoretical model and simulations are represented by a continuous and dashed lines resp.}}
\end{center}
\end{figure*}

\begin{figure*}[htbp]
\begin{center}
\includegraphics[width=\textwidth]{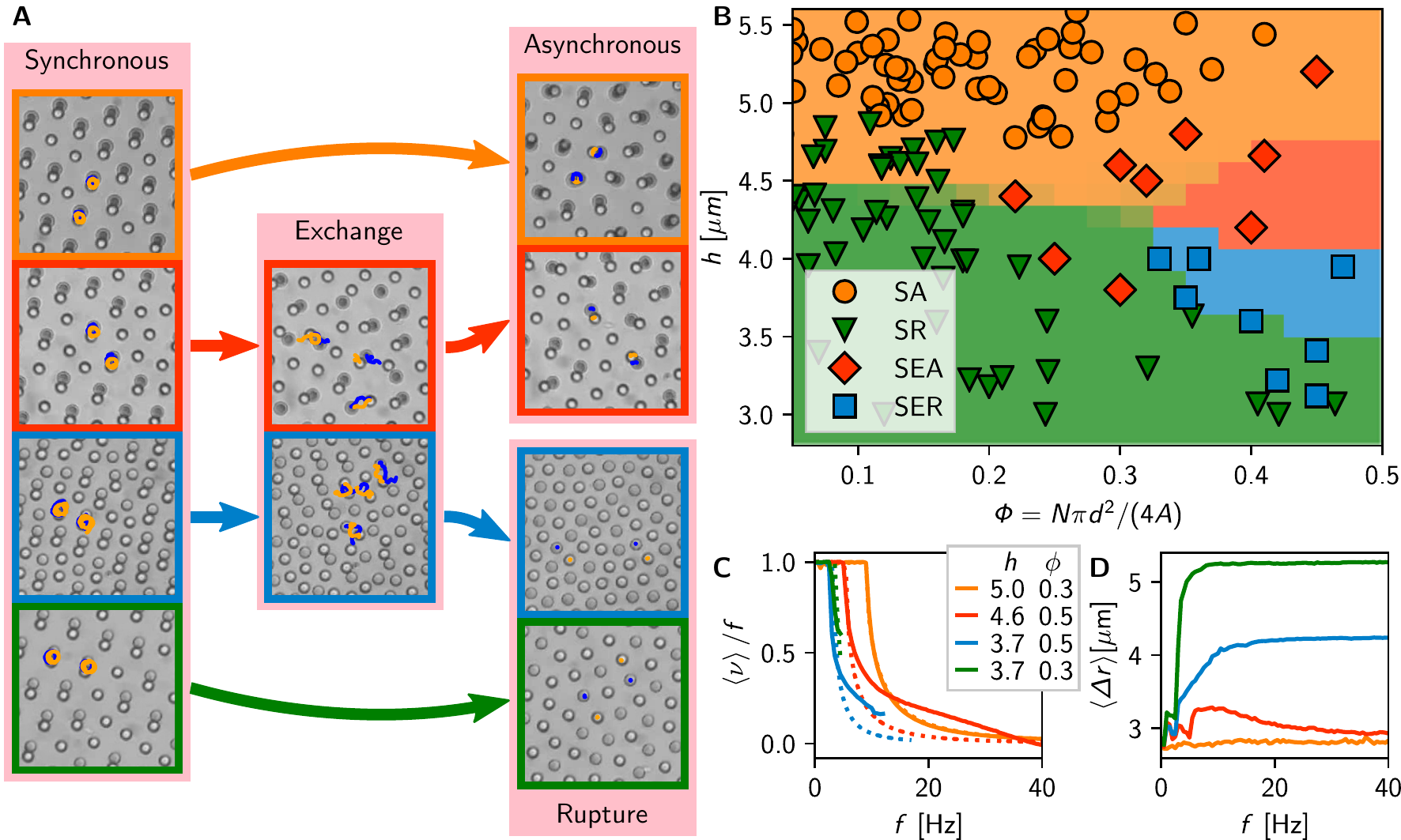}
\caption{\label{fig2}\textsf{\textbf{Dynamic colloidal states from experiments and numerical simulations.} 
(\textbf{A}) 
Images illustrating different transition paths observed for $B_0=7.28 \rm{mT}$, $\theta= 26.9^{\circ}$; all starting from a synchronous orbit at $f=1$Hz.
SA: Synchronous $\rightarrow$ Asynchronous rotations ($f=20$Hz), $h=5.1\rm{\mu m}$; 
SEA: Synchronous $\rightarrow$ Exchange of neighbors ($f=8$Hz) $\rightarrow$ Asynchronous ($f=25$Hz), $h=4.4\rm{\mu m}$;
SER: Synchronous $\rightarrow$ Exchange of neighbors ($f=3$Hz)  $\rightarrow$ Rupture ($f=14$Hz), $h=4.4\rm{\mu m}$; 
SR: Synchronous $\rightarrow$ Rupture ($f=9$Hz), $h=4.0\rm{\mu m}$. Movie S2 in ({\it 18\/}).
(\textbf{B}) Combined experiment (symbols) and simulation (colored regions) results of the location of the different transition paths in the $(\Phi,h)$ plane. (\textbf{C},\textbf{D})
From simulations: Average cyclotron frequency of the dimers $\nu$ (C) and 
nearest neighbor separation distance $\langle \Delta r \rangle$ (D)
versus driving frequency $f$. Both quantities can be used to classify the different transition paths observed.}}
\end{center}
\end{figure*}

\begin{figure*}[htbp]
\begin{center}
\includegraphics[width=\textwidth]{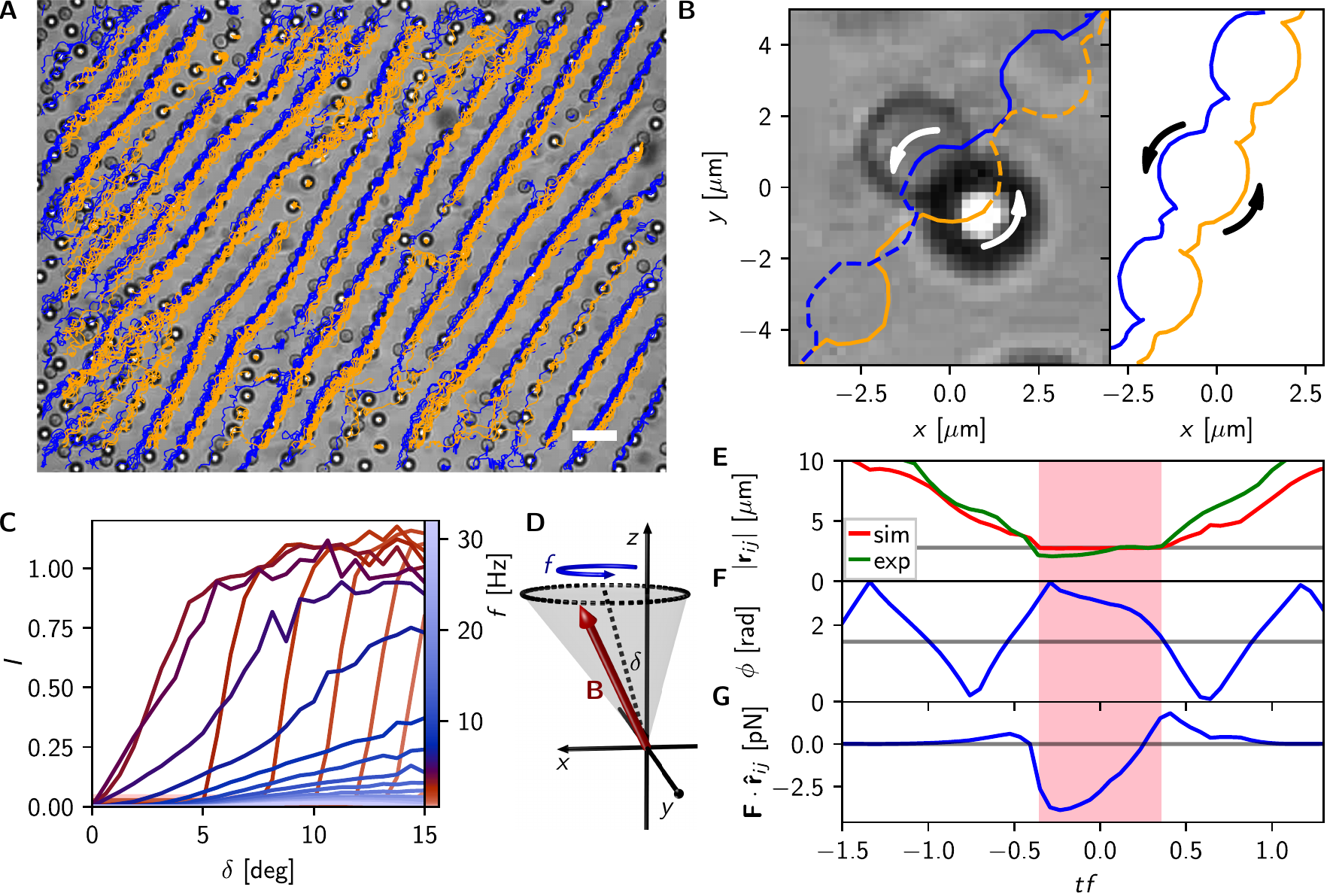}
\caption{\label{fig3}\textsf{\textbf{Colloidal edge current.} 
(\textbf{A}) Microscope image with superimposed trajectories of up (blue) and down (orange) particles driven by an off-axis precessing field with  $B_0=7.2 \rm{mT}$, $\theta= 26.9^{\circ}$ and $f=6$Hz in a cell of thickness $h=4 \rm{\mu m}$, Movie S3 in ({\it 18\/}). Scale bar is $10 \rm{\mu m}$.
(\textbf{B}) Enlargement of this image with only two superimposed particle trajectories from experiments (left) and simulation (right). 
(\textbf{C}) Intensity $I$ of the edge current versus shift angle $\delta$ for different values of the driving frequency (color bar) from simulations. 
High frequencies (blue lines) produce a smooth raise of $I$ to the maximum flow, where all up and down particles contribute to the current. For $f<4$Hz (red lines) we observe a sharp increase from zero current to the maximum value at a fixed $\delta$.
(\textbf{D}) Schematic showing a shifted conical field obtained by adding a zenithal angle $\delta$ which only rotates the static component of the magnetic field, 
$\bm{B}_{s} = \sin(\delta) \bm{e}_x + \cos(\delta) \bm{e}_z$. 
With this additional bias, the external field now reads as:
$\bm{B}(t) =
B_0 [\left(\sin(\theta)\cos(2\pi f t) + \cos(\theta) \sin(\delta) \right)\bm{e}_x + 
\sin(\theta)\sin(2\pi f t) \bm{e}_y + 
\cos(\theta)\cos(\delta) \bm{e}_z]$.
(\textbf{e}-\textbf{g}) Details on the exchange mechanism:  (\textbf{e}) inter-particle distance $\bm{r}_{ij}$ between two particles $(i,j)$, from experiments and simulations. Only from simulations:  
(\textbf{F}) phase lag angle between the dimer and the field $\phi$ and  (\textbf{G}) projection of the magnetic dipolar force $\bm{F} \cdot \bm{\hat r}_{ij}$ versus rescaled time $t$. The pink shaded region indicates the time window of the dimer formation, $r_{ij}=2d$. 
}}
\end{center}
\end{figure*}

\end{document}